\documentclass[a4paper,12pt]{article}
\usepackage{jcappub}
\usepackage{amsthm,graphicx,multirow}
\usepackage{epsfig}
\usepackage{latexsym, amssymb} 
\usepackage{amsmath}
\def\beq{\begin{equation}}
\def\eeq{\end{equation}}
\def\br{\begin{eqnarray}}
\def\er{\end{eqnarray}}
\def\benu{\begin{enumerate}}
\def\efnu{\end{enumerate}}

\begin{document}
\begin{center}
\title{See--Saw  Inflation / Dark Matter / Dark Energy / Baryogenesis:  \\
$ \begin{matrix}  & & \rm { \nu_L ~Dark ~ Energy} \\   & / & \\ \rm  GUT ~ \nu_R ~ \bf Inflation & & \end{matrix}$ }
\end{center}
\author[c,d]{George F. Smoot}

\affiliation[c]{Paris Centre for Cosmological Physics, APC (CNRS), Universit\' e Paris Diderot, \\
  Universit\'e Sorbonne Paris Cit\'e,  75013 France}
\affiliation[d]{Physics Department and Lawrence Berkeley National Laboratory, University of California, Berkeley, CA 94720, USA}
\emailAdd{gfsmoot@lbl.gov} 

\abstract 
{Motivated by BICEP2 results that imply gravitational waves are produced when the universe 
has an expansion energy of about $Mc^2  \approx 10^{14}$ GeV
and that a natural extension to the Standard Model of Particle physics is a right-handed neutrino
that would or could be at $m_{\nu_R}c^2  \approx 10^{14}$ GeV,
I propose here a See-Saw Inflation model which fits into the general class of models 
we have dubbed ``Wiggly Whipped Inflation"\cite{Hazra:2014WWI}.
The same scalar boson that stabilises the heavy right-handed neutrino mass then 
becomes the Inflaton whose potential is set by self-coupling and the heavy right-handed neutrino mass coupling.
Following this See-Saw Inflation one also finds that the seesaw mechanism provides for an after electroweak symmetry breaking offset in the Inflaton potential at the  near GUT scale vev by an amount set by the lightest neutrino and thus as a consequence produces a ``Dark Energy" at a scale set by the lightest left-handed neutrino mass $m_{\nu_L} \sim 10^{-3}$ eV.
Since these neutrinos are Majorana particles and violate lepton number conservation, 
they are the natural mediators of baryogenesis at the electroweak scale.
The residual heavy right-handed neutrinos work out to be Dark Matter with some fine tuning.
So an unaesthetic fine-tuned model can address the four remaining fundamental issues of cosmology 
by linking them to the neutrino sector.
The resurrected SO(10) GUT model provides a framework for the development of this scenario and allows a specific prediction for the Inflaton potential and a framework for fitting to neutrino as well as cosmological observables.
}

\maketitle
\section{Introduction}
The recent BICEP2 report~\cite{BICEP2:Detection,BICEP2:datasets} of signal consistent with the signal signature 
of primordial gravitation waves has consequences for the inflationary scenario: first is that gravitational waves are produced at an expansion energy of about $10^{14}$ GeV .
These are important consequences and their strength depends upon the confirmation and improvement of the BICEP2 results. 
We proceed under the assumption that the BICEP2 report is essentially correct for the purposes of this paper.

In the framework of the Inflation mechanism of primordial gravitational wave (GW) production, 
what BICEP2 has determined is  the expansion rate $H(k)$ for the scale corresponding to $\ell = 30$ to 100 and thus $k \sim 0.002 ~ {\rm Mpc}^{-1}$.  
Inflation tensor perturbations arise from quantum vacuum fluctuations,
whose amplitude is set by the de Sitter temperature $T_{dS} = H/2 \pi$ and thus the expansion rate $H(k)$.
The power spectrum of tensors (to be gravitational waves) is given by
\beq
P_h(k)  \equiv r(k) P_s(k) 
\approx\frac{2}{\pi^2} \left( \frac {H(k)} {m_{Pl}} \right)^2
\eeq
where $H(k)$ is the expansion rate when the $k$-mode exits the horizon and 
$M_{P} = \sqrt{\hbar c / G} = 1.22 \times 10^{19}$ GeV/c$^2$ is the Planck mass. 
In general we use the reduced Planck mass $m_{Pl} = \sqrt{\hbar c / 8 \pi G} = 2.435 \times10^{18}$ GeV/$c^2$.
Using the scalar normalization $P_s = 2.2 \times 10^{-9}$ at the pivot scale $k =$ 0.05 Mpc$^{-1}$ and scaling in $k$, 
more precisely 
\begin{eqnarray}
H_{k=0.002} &=& 0.99 \times 10^{-5} \left(\frac {r_{k = 0.002}} {0.2}\right)^{1/2} 5^{(0.96-n_s)} M_P  \\
&=& 4.97  \times 10^{-5} \left(\frac {r_{(k = 0.002}} {0.2}\right)^{1/2} 5^{(0.96-n_s)} m_{Pl}  ~~\approx  ~ 10^{14} {\rm GeV} .
\end{eqnarray}
where $r_{k = 0.002} $ is the tensor to scale ratio at $k = 0.002$ Mpc$^{-1}$ and $n_s$ is the scalar perturbation spectral index~\cite{Hazra:2014WWI}.
So, $H_{k=0.002}$ actually corresponds to an expansion energy of  $10^{14}$ GeV,
which sets the mass scale of the slow roll inflation potential and in the case of chaotic inflation
$M \sim 10^{14}$ GeV for its potential $V(\phi) = M^2 \phi^2$.

Now, with CMB temperature and E and B-mode polarization anisotropy spectra, 
we can determine the inflaton potential, its slope and, 
importantly for particle physicists, 
the effective inflaton mass without the need to know the underlying microscopic field (string, M-, etc.) theory.
It then becomes an important task to find a particle or pseudo particle with a mass around $10^{14}$ GeV.

The seesaw mechanism needs a right-handed neutrino with a mass of $m_{\nu_R}c^2  \approx 10^{14}$ GeV,
so it is natural to consider See-Saw Inflation based upon the very massive right-handed neutrino(s) and the Inflaton(s) being the scalar field(s) which gives the right-handed neutrino(s) their mass.
The relation to the general Wiggly Whipped Inflation model class is that there will be a transition from the dominant fourth power of the Inflaton field in the potential $V(\phi)$ to add a $ m_{\nu_R}^2  \phi^2$ term at the appropriate energy scale, e.g. GUT. 
When the electroweak or chiral symmetry breaking occurs and the seesaw mechanism is allowed to come into action, there is a very small added term, of order $ m_{\nu_L}^2  \phi^2$. 
This extra term is the right order to produce Dark Energy. 

The paper is organized as follows:
Section 2 is about the seesaw mechanism.
Section 3 is about See-Saw Inflation.
Section 4 is about See-Saw Dark Energy
Section 5 is about See-Saw Dark Matter
and then Section 6 is the Conclusion and Discussion.
\section{Seesaw Mechanism}~\label{sec:Seesaw}
For decades many people have noticed that neutrinos have such low masses, 
while all other particles, such as electrons and quarks,  are much heavier, 
with their masses relatively closely grouped.  
In the Standard Model particles get mass via the Higgs mechanism, 
so the question arises: 
Why, for example, should the electron neutrino be at least $10^5$ times lighter than the electron, up and down quarks?  
That is, why would the couplings of the neutrinos to the Higgs field be so much less by many orders of magnitude?

One proposed mechanism allows the possibility that after symmetry breaking, 
two types of neutrino exist, with one having zero mass (no Higgs coupling) 
and the other having (large) mass of the symmetry breaking scale gained though another source.
A small superpositions of these fields can result in light neutrinos (like those observed) and a still very heavy neutrino (of symmetry breaking scale, and as of yet unobserved).
This mechanism will result in a mixing, where  the Dirac mass $m_D$ is the geometric mean of the left and right Majorana masses, 
\beq
m_M^R \times m_M^L = m_D^2
\eeq
thus when the right-handed neutrino mass goes up the left-handed neutrino goes down and vice versa.
Hence the name the ``seesaw" mechanism.

The Lagrangian for the mass terms is
\beq
 L_{mass ~ terms} =  -\frac{1}{2}  \begin{bmatrix}\bar \nu_L & \bar \nu_R^c \end{bmatrix}  M \begin{bmatrix}\nu_L^c \\ \nu_R \end{bmatrix} + h. c.
~~~ {\rm where}  ~~~~
M =  \begin{bmatrix} m_M^L & m_D \\ m_D & m^R_M \end{bmatrix}   
\eeq
and $m_M$ is the Majorana mass and $m_D$ is the Dirac mass.
If no Dirac mass is the given end result of the mixing, then this becomes
\beq
 L_{mass ~ terms} =  -\frac{1}{2}  \begin{bmatrix}\bar \nu & \bar N \end{bmatrix}  \tilde M \begin{bmatrix}\nu \\ N \end{bmatrix} + h. c.
~~~ {\rm where}  ~~~~
\tilde M =  \begin{bmatrix} m_\nu & 0 \\ 0 & M \end{bmatrix}   
\eeq
Heuristically, the two basis vectors should be related to each other by a unitary matrix which is a rotation between the states to give the alignment with no residual Dirac mass.
In principle we would expect three right-handed neutrinos to correspond to the three flavours.

If we assume that this high mass right-handed neutrino exits and, what is more, that it is the mass that couples to the Inflaton to produce the Inflation we see in the horizon, then it needs to have a mass of about $10^{14}$ GeV
based upon the BICEP2 observations.
This in turn gives us a left-handed neutrino Majorana mass of about $10^{14}$ GeV. 
Table 1 presents some representative values for See-Saw.
\renewcommand{\arraystretch}{1.1}
\begin{table*}[!htb]
\begin{center}
\vspace{1pt}
\begin{tabular}{|c | c | c | c | }
\hline\hline
 \multicolumn{4}{|c|}{{\bf See-Saw Parameters}}  \\

\hline
 Flavor & electron ~ or ~1 & muon ~ or ~2 & tau ~ or ~3 \\
 \hline
 
 Mass (MeV) &  0.511 &  105.6 & 1776.8  \\
 \hline
 
 $\Delta(m_\nu^2)$ (eV$^2$) &  $\Delta M^2_{31} =  2.43  \times 10^{-3}$ & $\Delta M^2_{12} =  0.759 \times 10^{-4} $ & $\Delta M^2_{32} =  2.32 \times 10^{-3}$ \\
 \hline
 
 $m_{\nu_L} $ (eV) &   $\sim 0.0007$ &   $\sim 0.009 $ & $\sim$ 0.06 \\ 
 \hline
 
  $m_{\nu_R}$ (GeV )  for $m_\ell$ &  $3.6 \times 10^{5}$  & $1.2 \times 10^{9}$  & $5 \times 10^{10}$       \\
\hline

 $m_{\nu_R}$ (GeV); $\gamma$vev = 100 GeV & $ 1.4 \times 10^{16}$ & $ 10^{15}$  & $1.6 \times 10^{14}$ \\
 \hline
 
$\gamma$vev  (GeV);  $m_{\nu_R} = 10^{14}$ GeV &  8.5  &  30   &  80  \\

\hline\hline
\end{tabular}
\end{center}
\caption{~\label{tab:bestfits} Sample See-Saw parameters:
The last three rows assume: 
First the Dirac mass is about the associated lepton flavour mass,
which gives too low and too spread out results. 
Next we try the generally used Dirac mass as the Higgs vev times a neutrino coupling for the Dirac mass. 
This would be vev = 246 GeV  and so for estimate use $\gamma$vev = 100 GeV 
Finally working the problem backward assume that $m_{\nu_R}$ is about $10^{14}$ GeV as a fit to Inflation would suggest and find  $\gamma$vev and see that they are reasonable numbers. 
It is interesting to note that using a vev of 100 GeV one gets right-handed neutrino masses that correspond to critical energies: two GUT symmetry breakings and the Inflaton mass scale.
} 
\end{table*}

Thus we see that it is reasonable to expect the right-handed neutrino masses to be in the range of $10^{14} ~-~ 10^{16}$ GeV. 

However, our work is not quite done here. 
The flavor eigenstates are mixed by the weak interactions.
In the Standard Model of particle physics, the Cabibbo-Kobayashi-Maskawa matrix (CKM matrix, quark mixing matrix, sometimes also called KM matrix) is a unitary matrix which contains information on the strength of flavour-changing weak decays. 
The Pontecorvo-Maki-Nakagawa-Sakata matrix (PMNS matrix), Maki-Nakagawa-Sakata matrix (MNS matrix), lepton mixing matrix, or neutrino mixing matrix, is a unitary matrix, which contains information on the mixing of quantum states of leptons when they propagate freely relative to when they take part in the weak interactions.
\beq
 \begin{bmatrix}\nu_e \\ \nu_\mu  \\ \nu_\tau \end{bmatrix} = \begin{bmatrix}U_{e1} & U_{e2} & U_{e3} &\\U_{\mu 1} & U_{\mu 2} & U_{\mu 3} \\ U_{\tau 1} & U_{\tau 2} & U_{\tau 3}\end{bmatrix}  \begin{bmatrix}\nu_1 \\ \nu_2  \\ \nu_3 \end{bmatrix} 
\eeq
So we can expect that there will, as a result of the left-handed neutrino mixing, for the right-handed neutrinos to be also mixed as they are related by the seesaw matrix.
One can then think of one grand neutrino matrix
\beq
 \begin{bmatrix}\nu_e^R , \nu_e \\ \nu_\mu^R  ,  \nu_\mu  \\ \nu_\tau^R , \nu_\tau \end{bmatrix} = \begin{bmatrix}U_{e1} & U_{e2} & U_{e3} &\\U_{\mu 1} & U_{\mu 2} & U_{\mu 3} \\ U_{\tau 1} & U_{\tau 2} & U_{\tau 3}\end{bmatrix}  \begin{bmatrix} N_1 , \nu_1\\  N_2 , \nu_2 \\ N_3 , \nu_3 \end{bmatrix} 
\eeq

\subsection{Some Guidance from SO(10) Grand Unification} 
In SO(10) the unification of matter is even more complete than SU(5), since the irreducible spinor representation 16 contains both the $\mathbf{\overline{5}}$ and 10 of SU(5) and a right-handed neutrino, and thus the complete particle content of one generation of the extended standard model with neutrino masses. 

From the PDG\cite{PDG:2014} review of Grand Unified Theories we have: ``The boson matrix for SO(10) is found by taking the 15x15 matrix from the 10+5 representation of SU(5) and adding an extra row and column for the right handed neutrino. The bosons are found by adding a partner to each of the 20 charged bosons (2 right-handed W bosons, 6 massive charged gluons and 12 X/Y type bosons) and adding an extra heavy neutral Z-boson to make 5 neutral bosons in total. The boson matrix will have a boson or its new partner in each row and column. These pairs combine to create the familiar 16D Dirac spinor matrices of SO(10)."
They also provide us with the scalar fields needed to give the right-hand neutrino mass and be the Inflaton.

The minimal SO(10), does not require SUSY, but does provide for the seesaw mechanism.
Key issues are a home for the massive neutrinos being Majorana and thus not preserving $L$ quantum number and allowing for Baryogenesis through B-L. 
It contains a scalar field analog of the Higgs but that couples to the right-handed neutrino giving a natural vehicle to produce a high-mass ($\sim 10^{14} - 10^{15}$ GeV) right-handed neutrino.
The self coupling of the field plus the mass term $- (m_M^R)^2 \phi^2$ term in order to stablize the high right-handed neutrino mass.

What we have is the possibility of three singlet right-handed neutrinos and Higgs-like scalar fields to give them mass. 
All of this is in a symmetry group that reaches near the grand unification energy. 
There may well be multiple symmetry breakings in that high energy range as SO(10) has many ways to break down to smaller symmetries.

\section{See-Saw Inflation}
At early times we assume that the neutrino mass matrix has the form
\beq
 L_{mass ~ terms} =  -\frac{1}{2}  \begin{bmatrix}\bar \nu & \bar N \end{bmatrix}  \tilde M \begin{bmatrix}\nu \\ N \end{bmatrix} + h. c.
~~~ {\rm where}  ~~~~
\tilde M =  \begin{bmatrix} 0 & 0 \\ 0 & M_M^R \end{bmatrix}   
\eeq
That is to say, the coupling is to the right-handed neutrino.
So we are assuming that the relevant field only couples to right-handed neutrino and gives it a large mass.
This is the opposite of the weak interaction which couples only to the left-handed particles.
There is a symmetry here left to be understood, since for the Majorana neutrino is its own antiparticle. 

This gives us an additional quadratic term in the Inflaton potential in addition to the usual $\lambda \phi^4$
which is the coupling between the neutrino Majorana mass producing field and the right-handed neutrino field.
At high energies as the field drops below the phase transition one has
$$V_{eff}(\phi) =  \lambda \phi^4 \rightarrow  -\gamma m_{\nu_R}^2 \phi^2 + \Delta V + \lambda \phi^4  $$
where $\Delta V$ represents the corrections and the coupling constant $\gamma$ should be order unity
and the self-coupling coefficient for the field is $\lambda$.

The effective potential can be Taylor series expanded around the minimum.
\begin{eqnarray}
V_{eff}(\phi) &=&  -\gamma m_{\nu_R}^2 \phi^2 + \Delta V + \lambda \phi^4  \\
V_\phi &=&  -2 \gamma m_{\nu_R}^2 \phi + 4  \lambda \phi^3   ~~~~~~\phi_0 = m_{\nu_R} \sqrt{\gamma/2 \lambda} \\
V_{\phi \phi} &=& -2 \gamma m_{\nu_R}^2 + 12  \lambda \phi^2  \\
V_{\phi \phi \phi} &=& 24 \lambda \phi \\
V_{\phi \phi \phi \phi} &=&24 \lambda
\end{eqnarray}
Thus expanding about $\phi_0$ and assuming the correction $\Delta V$ brings the effective potential to zero there, 
as required by the SO(10), SU(5), SU(2) symmetries,
we have
\beq
V_{eff}(\phi)  =4 \gamma m_{\nu_R}^2 \Delta \phi^2 + 4  \lambda \phi_0 \Delta \phi^3 + \lambda \Delta \phi^4
\eeq
where $\Delta \phi = \phi - \phi_0 = \phi - m_{\nu_R} \sqrt{\gamma/2 \lambda}$.
The primary point being that we have terms that are quadratic, cubic, and quartic in $\Delta \phi$
and will be dominated be the quadratic potential when $\Delta \phi < M_{Pl}$.

Going back to the full form of the potential with the assumption of our symmetry forcing the minimum energy to zero with the corrections we have $\Delta V = \gamma^2 m_{\nu_R}^4 / 4 \lambda$ and thus the full form of the potential would become
\beq
V_{eff}(\phi) =    -\gamma m_{\nu_R}^2 (\phi^2 -\gamma m_{\nu_R}^2 / 4 \lambda)  + \lambda \phi^4 
\eeq
It would be natural to expect the self-coupling coefficient $\lambda$ would be related to the GUT and Planck scales:
$$\lambda = \left( \frac{ M_{GUT}} { M_{P}}\right)^4 \sim \left( \frac{ 2 \times 10^{16} ~ {\rm GeV}} { 1.2 \times 10^{19}~ {\rm GeV}}\right)^4
\sim 10^{-11}  {~\rm or ?} ~ \sim \left( \frac{ 2 \times 10^{16} ~ {\rm GeV}} { 2.435 \times 10^{18}~ {\rm GeV}}\right)^4
\sim 10^{-8}  $$
Let us consider for example the case where $\lambda = 10^{-10} $ and $\gamma m_{\nu_R}^2 = 10^{-11} M_{Pl}^2$,
then $\phi_0 = 0.22 M_{Pl}$.

Presumably, the effect producing the high right-handed neutrino mass turns on at the GUT symmetry breaking
and adds the mass and correction term to the potential. 
If we want that to be a second order transition, that is continuous but with a break in slope,
 then we have the requirement that at that point
$ \phi^2 -\gamma m_{\nu_R}^2 / 4 \lambda = 0$ or $\phi^2 = \gamma m_{\nu_R}^2 / 4 \lambda$ leaving the potential at $\lambda \phi^4$ which reaches the GUT energy at $\phi = M_{Pl}$, if the relation for $\lambda$ is correct 
and then the consequence would be that $\gamma m_{\nu_R}^2  = 4 \lambda M_{Pl}^2$ or that 
$\sqrt{\gamma} m_{\nu_R}$ is the GUT scale scaled down by the ratio of the GUT scale to the Planck scale
or roughly $10^{-2} $ of the GUT scale or about $2 \times 10^{14}$ GeV.
At that point we end up with a break in slope from $4 \lambda \phi^3$ to 
$-2 \gamma m_{\nu_R}^2 \phi + 4 \lambda \phi^3$ and putting in characteristic numbers this would roughly 
be from quartic potential to a quadratic but with a transition, i.e. what is needed to make BICEP2 and Planck results more compatible. 

One also has the option of not making the transition continuous by having a step in potential or by having more than one symmetry breaking as is natural in SO(10). 
Thus there is a range of parameter space in the See-Saw Inflation potential to both produce the low-$k$ scalar suppression to reconcile BICEP2 and Planck observations but also produce wiggles in the scalar perturbation spectrum that would fit known features in the CMB observations and predict features in the processed matter power spectrum. 
See-Saw Inflation is a fairly direct variant of our ``Wiggly Whipped Inflation"\cite{Hazra:2014WWI} class of models
and can be expected to exhibit similar features.

There is an issue of there being three right-handed neutrinos all with masses in this neighbourhood.
The question is do they all couple to the same field and thus have the same mass or do they couple to a triplet of fields and can have slightly separate masses. 
If they couple to one field then presumably the $\gamma m_{\nu_R}^2 = 3 m_{\nu_R}^2$. 
If they couple to separate fields then there might be three separate Inflatons with different vevs and therefore fit into a bumpy road of inflation. 
That is one inflation followed and interrupted by another.
Now one feature of SO(10) is that it works without supersymmetry and thus without out the need for an Inflatino.
For this scenario description I assume that there is only one Inflaton to worry about and that the symmetry requires the masses and fields to match conveniently. It should be straightforward to include all three.

In the framework of SO(10) coming down from the Planck scale one may well have a resonance at the SO(10) GUT energy scale which then is slightly above the SU(5) symmetry breaking scale.
An issue is that above the SO(10) scale the coupling may decrease toward the Planck scale and symmetries might become more restrictive. 

Thus the turn on of the high-mass right-handed neutrino(s) might be very or relatively abrupt
and thus not only add the Inflaton field $\phi$ squared term and potential correction to zero to the potential but also do it fairly abruptly.
Thus the possible wiggles and relationship to Whiggly Whipped Inflation which involve a rapid transition in the slope of the inflation potential at an effective energy of about $(10^{16}~GeV)^4$.
This fits the data and if the transition has the appropriate features, explains additional features in the CMB
temperature power spectrum.

As soon as the Electroweak /  Chiral symmetry is broken and the see-saw mechanism comes into play,
the right-handed neutrino donates a small mass to the left-handed neutrino by going to the eigenstate given by the see-saw mechanism.

\section{See-Saw Dark Matter}
While I was rushing to submit this concept paper made bricolage\footnote{In the practical arts and the fine arts, bricolage (French for "tinkering") is the construction or creation of a work from a diverse range of things that happen to be available. Now has a somewhat disparaging accent in French.} style, my colleague Ivan Debono gave the paper a quick pass and commented ``Why didn't you also explain the Dark Matter too?"
I thought it a flippant remark but did know that there would be some residual heavy right-handed neutrinos but 
did not pause to take the time to calculate if the thermal relict calculation would leave in place the appropriate number
at the time of submission. 
If it had the weak interaction as its cross section then it would be too rare. 
The rush was also why there were so few references.

As soon as the paper was submitted, I received the usual number of emails from people saying that  effectively ``you should have cited my earlier work\cite{usual:2014}" even though most do not have much to do directly with the paper.
However, one of these emails from Qaisar Shafi pointed to his article 
``Inflation and majoron dark matter in the seesaw mechanism" by
Sofiane M. Boucenna, Stefano Morisi, Qaisar Shafi, Jose W.F. Valle\cite{Boucenna:2014}
in which they ``propose that inflation and dark matter have a common origin, connected to the neutrino mass generation scheme."  
In this they generate a keV energy scale neutrino and also discuss the B-L and lepton number violation as a way to baryogenesis.

So I felt compelled to put in a discussion of See-Saw Dark Matter in the context of this work.

The issue of ultra-massive dark matter particles has been studied before.
We go back to the study of WIMPZILLAS.
For a reheating temperature of the order of 100 GeV, 
the present abundance of WIMPZILLAS with mass $M_X \approx 10^{14}$ GeV is given 
by  $\Omega_X \sim1$ if $\epsilon \sim 10^{-10}$. 
This small fraction corresponds to $< X^2 > \sim 10^{-12} M_{Pl}^2$ at the end of the preheating stage, 
a value naturally achieved for WIMPZILLA mass in the GUT range\cite{Khlebnikov:1997}. 
The creation of WIMPZILLAS through preheating and, therefore, the prediction of the present value of $\Omega_X$, 
is very model dependent.
However, this shows that there are mechanisms for producing the correct residual density of 
heavy right-handed neutrinos to make the Dark Matter.

Massive right-handed neutrinos are in copious abundance at the GUT scale where they will share in the energy density  thermal equilibrium state of about an energy density of $m_{\nu^R}^4 \approx (10^{25} eV)^4$.
Then we have inflation which now only goes on 50 $e$-folds followed by another approximately $10^{25}$ 
in more conventional expansion. 
This expansion reduces the number density and thus the energy density of the right-handed neutrinos by 
the ratio of the scale factors cubed or about $(10^{50})^3$ to $(10^{-12.5} eV)^4$.
This is the same way that Inflation and expansion make magnetic monopoles scarce.


Now to consider the effect of the seesaw mechanism:
We start with the mass matrix early in the universe 
where the Lagrangian for the mass terms is
\beq
 L_{mass ~ terms} =  -\frac{1}{2}  \begin{bmatrix}\bar \nu_L & \bar \nu_R^c \end{bmatrix}  M \begin{bmatrix}\nu_L^c \\ \nu_R \end{bmatrix} + h. c.
~~~ {\rm where}  ~~~~
M =  \begin{bmatrix}  0 & 0 \\ 0 & m^R_M \end{bmatrix}   
\eeq
At electroweak symmetry breaking we anticipate that the Dirac mass term appears
\beq
 M_{early} =  \begin{bmatrix}  0 & 0 \\ 0 & m^R_M \end{bmatrix}   \rightarrow
M_{electroweak} =  \begin{bmatrix} 0 & m_D \\ m_D & m^R_M \end{bmatrix}   \rightarrow 
M_{seesaw} =  \begin{bmatrix} m_M^L & 0 \\ 0 & m^R_M \end{bmatrix}   
\eeq
presumably via a unitary operators.
An example would be  $U_{1} = \begin{bmatrix} cos(\theta) & 0 \\ 0 &sin(\theta) \end{bmatrix}$
giving 
\beq
M_{electroweak} =  \begin{bmatrix} 0 & m_D \\ m_D & m^R_M \end{bmatrix}    = U M_{early}
= \begin{bmatrix} cos(\theta) & 0 \\ 0 &sin(\theta) \end{bmatrix}  \begin{bmatrix}  0 & 0 \\ 0 & m^R_M \end{bmatrix}
=   \begin{bmatrix}  0 & sin(\theta) m^R_M  \\ sin(\theta) m^R_M  & cos(\theta)m^R_M \end{bmatrix}
\eeq
which gives us the relation $sin(\theta) = m_D / m^R_M$ and we anticipate that $m_D \sim 10^2$ GeV 
and that $m^R_M \sim 10^{14}$ GeV so that $sin(\theta) \sim 10{-12}$.
We note that there could be phases here.
In the normal course of thermal universe expansion the Dirac neutrinos would freeze out at a relative density,
in what was called the ``WIMP miracle", at the level to produce about $\Omega_{\nu s} \sim 0.25$.
Viola! We have the Dark Matter. 
What is it today? Why don't we see these 100 GeV Dirac neutrinos in our detectors.
Well our handy seesaw mechanism continues to operate (unitarily $U_2$) to turn that component into left and right-handed Majorana neutrinos.
\begin{eqnarray}
M_{seesaw} &=&  \begin{bmatrix} m_M^L & 0 \\ 0 & m^R_M \end{bmatrix} = U_2 M_{electroweak}  =
\begin{bmatrix} cos(\theta_2) & 0 \\ 0 & -sin(\theta_2) \end{bmatrix} \begin{bmatrix} 0 & m_D \\ m_D & m^R_M \end{bmatrix}  \\
&=& \begin{bmatrix} 0  & sin(\theta_2) \\ -cos(\theta_2) & 0 \end{bmatrix}  \begin{bmatrix}  0 & sin(\theta) m^R_M  \\ sin(\theta) m^R_M  & cos(\theta) m^R_M \end{bmatrix} \\
&=& \begin{bmatrix}  sin(\theta_2) sin(\theta) m^R_M   & \left( -cos(\theta_2) sin(\theta)+ sin(\theta_2) cos(\theta) \right) m^R_M  \\  0 & -cos(\theta_2) cos(\theta)m^R_M \end{bmatrix} \\
& = &  \begin{bmatrix} sin^2(\theta) m_R^M & 0 \\ 0 & - cos^2(\theta) m^R_M \end{bmatrix} 
~~~{\rm for}~~~  \theta_2 = \theta
\end{eqnarray}
Giving the usual $ m_M^L = sin^2(\theta) m^R_M = m_D^2/ m^R_M$ seesaw relationship.

The Dark Matter is simply the rotated combination of left-handed and right-handed massive neutrinos.
There will be only of the order one $10^{14}$ GeV right-handed massive neutrinos in a cube a 100 km on the side and they are currently moving slowly - orbital speeds or about $10^{-3} c$.
Note that the right-handed neutrinos being in singlets do not interact much except with the Inflaton fields 
and with gravity.
However, if there was an interaction they could rotate back to the Dirac form; 
but the number entering the detector is roughly $10^{-12}$ or  less of the rate of standard 10 to 100 GeV range WIMPS
and thus they would escape detection at the current level of direct detection sensitivity.
This may just be in range for the CERN experimental searches once the LHC turns on again in January 2015
as long as the trigger to a jet or gamma and a large missing component. 
This for the diagram such as $p +  \bar p \rightarrow  {\rm jet ~ or} ~\gamma + {\rm ~ off ~ mass ~ shell}~ Z $
and the off mass shell $Z \rightarrow  \nu_D + \bar \nu_D$.

\section{See-Saw Dark Energy}

One can not help but note that Inflation happens on a very high energy scale while the Dark Energy causing the current acceleration of the Universe's expansion is on a very low energy scale.
Consider the ratio of the energy scale of Inflation to the energy scale of its counter part Dark Energy
and compare it to the energy scale ratio for the neutrino masses
\beq
\frac{E_{DE} }{E_{Inf}} \sim  \frac{2 \times 10^{-3} ~ {\rm eV}}{2 \times 10^{16} ~ {\rm GeV}} \sim 10^{-28} 
~~~{\rm vs.} ~~~   \frac{m_{\nu^L}} {m_{\nu^R}} = \frac{1 \times10^{-3} ~ {\rm eV}}{2 \times 10^{14} ~ {\rm GeV}} \sim 10^{-26}
\eeq
The same seesaw mechanism that couples the right-handed massive neutrinos to the Inflaton also
couples the later left-handed neutrinos to the 
triplet of Inflatons that give the right-handed neutrinos their masses.
This is because they become  Majorana and their mass comes from the slight rotation of the right-handed massive neutrino and thus the slight rotation of the Inflaton. 
The potential changes from
\beq
V(\phi) = -{m^R_M} ^2 \phi^2 + \lambda \phi^4 \rightarrow  -(cos(\theta)^4 + sin(\theta)^4) {m^R_M} ^2 \phi^2 +  \lambda \phi^4
\eeq
This change in coupling/potential can introduce a potential level shift above the zero point energy of the condensate through two different routes, displacement of the field value from the zero point and coupling to neutrinos in a thermal bath.

Expanding 
\begin{eqnarray}
V(\phi) &=&  -(cos(\theta)^4 + sin(\theta)^4) {m^R_M} ^2 \phi^2 +  \lambda \phi^4 \\
&=& -((1 - sin(\theta)^2)^2 + sin(\theta)^2)  {m^R_M} ^2 \phi^2 +  \lambda \phi^4 \\
&=& -  {m^R_M} ^2 \phi^2 +  \lambda \phi^4  - 2 sind(\theta)^2 {m^R_M} ^2 \phi^2 \\
&=& -  {m^R_M} ^2 \phi^2 +  \lambda \phi^4  - 2 m_{\nu^L}^2 \phi^2
\end{eqnarray}
Meaning we recover the original potential plus a term $\Delta V(\phi) =  - 2 m_{\nu^L}^2 \phi^2$.
Thus we get a shift in the vev$_{\phi}$ from its original field value of $\phi_0 = m_{\nu^R} / \sqrt{2 \lambda}$
to a value shifted by  $\Delta \phi_0 = - m_{\nu^L} / \sqrt{2 \lambda}$.
One assumes that the field $\phi$ does not change instantly to this new minimum,
 but the field is in a shifted potential 
by an amount $\Delta V(\phi)|_{\phi_0} = - 2 m_{\nu^L}^2  \Delta \phi_0^2 \sim  m_{\nu^L}^4/ \lambda$ from the new minimum.

 The $V(\phi)$ potential is very flat at the vev (minimum) and,  depending upon how the light left-handed neutrino mass is turned on,  could move the minimum in the potential by an amount $\Delta \phi_0^2 = m_{\nu^L}^2 / 2 \lambda$ 
from true new $\phi_0^2 = (\gamma m_{\nu^R}^2 + m_{\nu^L}^2) / 2 \lambda$. 
The right-handed neutrino may well lose just enough mass (assuming $\gamma = 1$), if the transformation is not unitary, if the affect is simply a rotation to leave the minimum unchanged. 
If not the change in potential is
$\Delta V(\phi)|_{\phi_0} \sim  m_{\nu^L}^4 \times 4 \gamma m_{\nu^R}^2 / 4 \lambda^2 $ 
where the last part comes from the first term of the Taylor series expansion equation 3.7.
One expects that $\gamma m_{\nu^R}^2 /  \lambda \sim 1$
leaving the change in potential level of the field as roughly $\Delta V(\phi)|_{\phi_0} \sim  m_{\nu^L}^4 / \lambda$.
Here we get the energy scale of the lightest left-handed neutrino but enhanced by a factor 
of $1/\lambda$ with is roughly 11 orders of magnitude higher than one would want.
This assumes that the field $\phi$ was at its GUT vev (just the right-handed neutrino) and did not change instantly when the light left-handed neutrino turned on at the electroweak symmetry breaking
and thus has to slow roll from the old minimum to the new one.
This estimation give the same result as the simple unitary transformation approach.

However, since it appears there at the electroweak scale it slowly rolls down toward the new minimum
at a rate given by  $\dot \Delta \phi \sim V_\phi / 3 H$
where $H$ is not determined by the field $\phi$ or $\Delta \phi$ during this epoch 
but by other energy content in the universe and has only become important in this regard recently.
It would have come down by order $10^{-11}$ to $10^{-15}$ to be present residual
of order  $\Delta V(\phi) \sim ( 1 ~ {\rm to}~ 10^{-3} ) \times m_{\nu^L}^4 $ and then continues to roll slowly toward the minimum 
which is still zero according to whatever symmetry protects the vacuum energy.
However, one can estimate that the slow roll will be very slow and that it is practically frozen at the perturbed value.

We must remember that our field $\phi$ is a quantum field and that it will actually be a two component system.
The major portion is the bose condensate, $\phi_c$ and the much lesser portion is the ``normal" meaning a thermal fluid or the displaced fluid. 
I.e. $\phi$ is a superposition of states of the condensate with a thermal or displaced component $\phi*$ with an effective energy of the order of the displacement potential or the thermal bath from the GUT and the electroweak scales.
The energy in the $\phi*$ component will be shed by a process one could call ``betaless double neutrino decay"
\beq
\phi* \rightarrow \phi_c + 2 \nu_M^L + K.E. + 0 \beta
\eeq
The energy balance for this allowed decay would leave one with
\beq
E_{DE} = 2 m_{ \nu_M^L} c^2 + 3k_BT_{ \nu_M^L} + 2 \epsilon_F
\eeq
Turning this around one has a formula for the lightest left-handed Majorana neutrino mass
\beq
 m_{ \nu_M^L} c^2 = \frac{1}{2}E_{DE} - \frac{3}{2}k_B T_{ \nu_M^L} - \epsilon_F
 \eeq
Appropriate values give
$  m_{ \nu_M^L} c^2 = 10^{-3} - 3 \times 10^{-4} - ? \sim 0.7 \times 10^{-3}$ eV,
where have chosen $T_{ \nu_M^L} \sim 2.5 K$ and ignored the Fermi degeneracy energy for the neutrinos.
This explains the entry in Table one: See-Saw Parameters for the lightest mass.
Thus we not only find the lightest  left-handed Majorana neutrino mass to be roughly $10^{-3}$ eV from the 
Dark Energy energy density, we also find that the $\phi$ field will continue to adjust itself 
to match the neutrino temperature until that falls well below the neutrino mass energy.
That is the See-Saw field energy decreases as the temperature does, 
behaving like radiation plus a smaller vacuum energy, 
until about a redshift of one or two
and then flattened out to a level given by twice the lightest left-handed Majorana neutrino mass, 
behaving like vacuum energy at that point.

Now the energy balance is in terms of an energy density and thus the energy density must be at this level 
inside the Compton wavelength volume about $(h/mc)^3$ of the neutrinos.  
For the level of $m_{\nu_1} c^2 \sim 0.7 \times 10^{-3}$ the Compton wavelength is about a mm.
The energy density $\epsilon \sim m_{\nu_1} c^2 m_{\nu_1}^3 c^3 / h^3 \rightarrow m_{\nu_1}^4$.

Current analysis of the parameters of the Dark Energy may not have included the thermal portion
 simply by adjusting the expansion rate slightly and fitted to the form
$E_{DE} = 2 m_{ \nu_M^L} c^2 + 2 \epsilon_F $ 
in which case $m_{ \nu_M^L} c^2 \sim 10^{-3}$ eV as $2\epsilon_F$ is less than a tenth that.

One might be concerned that the heavier left-handed neutrinos would also couple moving the energy scale up by two orders of magnitude. 
However, since the light neutrinos, $\nu_1$, $\nu_2$, and $\nu_3$ are able to oscillate into each other, 
the vacuum can decay to the lowest energy states on a time scale short compared to the age of the universe.

The only coupling the Inflaton field $\phi$ has is to Majorana particles which would be 
the right-hand neutrino, the left-hand neutrino Majorana components, and the photon.
Like the Higgs couples only to Dirac particles with a coupling given by the mass,
the $\phi$ coupling coefficient is simply the square mass of the particle.
Thus the coupling to the photon is zero to first order.
I did not think of the higher order but there could be decay to two neutrinos which then make a loop to create a virtual Z intermediate vector boson and the Z decays to three photons. (The Z must go to one or three photons and if want zero net momentum, then it is three.) 
However, I should have as the Higgs has zero coupling to the photon (apparently only particle with no coupling to either)  but manages a decay fraction to two photons at about $2 \times 10^{-3}$ level.
I have not yet calculated the life-time given the weak interaction and the low energy - Z is way off mass shell and the phase space is small - I expect this to be much longer than the age of the universe.

Now for a rough calculation
the rough decay rate for $\phi* \rightarrow 0\nu + \gamma + \gamma+  \gamma$ to three gammas is given by
\beq
(T^{0 \nu}_{1/2})^{-1} = G^{0 \nu} |M^{0 \nu}|^2 < m_{\gamma  \gamma \gamma}>^2
\eeq 
where $T^{0 \nu}_{1/2}$ is the half life, $G^{0 \nu} $ is the phase space and 
$< m_{\gamma \gamma \gamma}>^2 = |\sum_i U_{e i}^2 ,_{\nu_i} |^2$ is the square of the effective Majorana mass.
The phase space $G^{0 \nu} $ is roughly $1/8 \pi$ and the matrix element 
$ |M^{0 \nu}|^2 \approx g_{weak}^2 / ( p^2 - M_z^2) \sim ~{\rm very ~ long}.  $

As a last ugly resort one could indicate that the last symmetry breaking also end the symmetry that makes the zero-point energy at zero and the natural break would we set by about twice the neutrino mass and that gives
the zero point offset of the inflation zero-point energy.

Thus the coupling at the Electroweak scale is one that allows the Inflaton field to relax significantly into neutrinos which then generate other particles in thermal equilibrium.  
The Inflaton is left in a thermal equilibrium state with the bath at that time. 
As the universe expands the temperatures decrease until one gets to the floor set by the lightest neutrino mass and the Fermi energy which also decreases as the universe expands so the real floor is twice the lightest neutrino mass.

\section{See-Saw Baryogenesis}

It is well known that simple Leptogenesis and thus through B-L Baryogenesis is possible, 
if reheating of the universe is $T_{reheat} \sim > 100$ GeV  
({Kuzmin, Rubakov, and  Shaposhnikov 1985}).
\beq
N^{fin}_{B-L} = \sum_i \epsilon_i \kappa^{fin}_i \approx  \epsilon_1 \kappa^{fin}_1
\eeq
wheere 
\beq
\epsilon_i \equiv  - \frac{\Gamma_i - \bar \Gamma_i}{\Gamma_i + \bar \Gamma_i}
\eeq
is the lepton asymmetry generated to a heavy particle decay process and is set by CP violation 
(Fikugita and Yanagida 1986)
and $\kappa_i$ is the decay parameter 
\beq
\kappa_1 \equiv \frac{\Gamma_{N_1}}{H(T = M_1)} \sim \frac{m_{sol,atm}}{m_x \sim 10^{-3} ~{\rm eV}} 
\sim  10 ~{\rm to} ~ 50
\eeq

There is an upper bound on $\epsilon_1$,  $\bar \epsilon(M_1) \approx 10^{-6} \left( \frac{M_1}{10^{10 }~ {\rm GeV}} \right) $
(Davidson and Ibarra 2002; BuchmŸller, DeBari and,PlŸmacher 2003; Hambye et al 2004 DeBari2005 ).
This in turn sets a strong lower bound on the mass of the lightest right handed neutrino  
of about $M_1 > 3 \times 10^9$ GeV. 
Fortunately this is fine in the See-Saw model and we do not run into the bound that it must be lower than about 
$2 \times 10^{15}$ GeV. 
This range works well for our See-Saw model but one has to be careful about which favour or 1 is which.

Thus the See-Saw model easily incorporates Baryogenesis through leptogenesis at the electroweak scale
and we see that the early universe knows about the neutrino masses during leptogenesis.

\section{Discussion and Conclusions}
We have a natural See-Saw Inflationary scenario based upon having heavy ($10^{14}$ GeV) right-handed neutrinos to explain the observed light left-handed neutrinos. 
There needs to be a scalar field to produce the heavy right-handed neutrino mass and it is a natural source for inflation - See-Saw Inflation.

This leads to plausible and generally physically possible though perhaps fine-tuned mechanisms to tie 
the neutrino sector to the four major fundamental issues in cosmology: Inflation, Dark Matter, Dark Energy,
and Baryogenesis.

The weakest argument presented her is for Dark Energy.
It is not unreasonable to find that this inflation gets back into action 
when it is perturbed by the later symmetry breaking 
allowing the left-handed neutrinos to gain a very low Majorana mass,
and then make a very slow roll and reasonably rapid decay 
to the new minimum producing the apparent Dark Energy accelerating the universe.
Thus the seesaw mechanism completes its work.

In one simple incarnation there are three right-handed neutrinos and related fields that correspond to energy levels of two GUT symmetry breakings, e.g. SO(10), and the last big Inflation.
The lightest right-handed neutrino acting with its field produces the last high-scale inflation period
and then the lightest left-handed neutrino acting with its field produces the late-time accelerating universe.

It is interesting to note that the lowly left-handed neutrino and its high-borne right-handed neutrino partner 
appear to have a big role in the destiny of the universe and that measurements of the neutrino properties reflect on parameters both at the GUT scale and at the lowest energy scale. 
In particular, it is important to determine:

1) Are these Majorana neutrinos?  So I say to my Cuore colleagues go to it, as I know have a personal interest beyond being involved via (former) graduate student Michele Dolinski, post doc Tom Gutier in Cuoricino and my colleagues in Berkeley. Let's see some good neutrinoless double-beta decay.

2) Measuring the neutrino mass spectrum as these feed directly into the fits for the right-handed neutrino mass and the inflation potential. 
There will undoubtedly be joint fits between the neutrino data and the large scale structure and CMB data to make a global fit to the right-hand neutrino mass and the inflation self coupling.

The large scale structure observations - e.g. galaxy and quasar-Lyman alpha surveys - may have interesting things to say not only about levels and coefficients but also about any structure in the potential or any splitting or right-hand neutrino masses.

Theorists have their work cut out in continuing the resurrection of SO(10) and making a more coherent and exhaustive treatment of the neutrino and scalar sectors and the links to observables.

In discussions Alexei Starobinsky said to me: 
``A good way to show broad public why it can be misleading to think
about the energy density scale (or a potential scale) $E$ defined as
$\rho c^2= E^4/(\hbar c)^3 $ as a characterictic energy scale of a substance
involved is to consider water, or even our body, with $\rho = 1~ g/cm^3$. 
Then $E\approx 45$ keV, and it is clear that no physical, chemical or biological process with water has such a characteristic energy scale.''
My response is that is correct, there are no biological of chemical processes ongoing in the human body
at the level of 45 keV.  (except for an occasional radioactive decay)
However, to get that energy density nuclear physicists would tell us that there are nuclear potentials involved at the MeV to tens of MeV level. 
A high energy physicist would say yes and there mass generating potentials up to the 1 GeV energy scale 
to get the approximate 1 GeV masses for the protons and neutrons inside the nuclei.
Thus an energy density at the level of $(10^{14}  ~ {\rm to} ~10^{16} {\rm GeV})^4$ level implies that there are potentials in action at that or higher scales. 
Now from BICEP2 we have no evidence that things are taking place at higher than the $10^{14}$ GeV scale.
In the context of Inflation, one finds $10^{16}$ GeV coming up naturally as the energy density in the Inflaton field potential. The natural ones are there and above are the GUT scale around $2 \times 10^{16}$ GeV
and the Planck scale of $2 \times 10^{18}$ GeV.
So there is cause for excitement even though we are only now seeing evidence of particles interacting at the
10$^{14}$ GeV level.

\section*{Acknowledgments}
G.F.S. acknowledges the financial support of the UnivEarthS Labex program at Universit\' e Sorbonne Paris Cit\' e (ANR-10-LABX-0023 and ANR-11-IDEX-0005-02). 
I thank my ``Wiggly Whipped Inflation"\cite{Hazra:2014WWI}  coauthors Dhiraj Hazra, Arman Shaffieloo,
and Alexei Starobinsky for focusing my attention on this topic so fiercely and making clear the need for both a 
$10^{14}$ GeV mass scale and a GUT-scale potential so that I was strongly motivated to find a combination that would work.  Soon after I alighted on the see-saw right handed neutrinos, Pierre Binetruy pointed me away from the standard Higgs to look for the scalars to be the Inflaton. Frederico Piazza also echoed that need. This lead me to look to SO(10) as a natural framework for the see-saw mechanism, heavy massive right-handed neutrinos and appropriately coupled scalar fields that could be the Inflaton.  Eric Linder provided insight and corrections and pointed out the weakness in the See-Saw Dark Energy arguments.
 


\begin{thebibliography}{99}
\bibitem{usual:2014}
Long list of moderately to immoderately not related preprints
To be replaced by a list of references and a rating of the relevance 
and the effectiveness of the claiming author.
\end{thebibliography}
\end{document}